\documentstyle[seceq,epsf]{ptptex}

\def\cm{\,{\rm cm}}
\def\erg{\,{\rm erg}}
\def\sec{\,{\rm sec}}
\def\F{\,{\cal F}}

\title{
Pulsar Kick and Asymmetric Iron Velocity Distribution in SN 1987A
}

\author{
K. {\sc Kohri}$^{1}$
and S. {\sc Nagataki}$^{2}$
}

\inst{
$^1$Research Center for the Early
Universe, Faculty of Science, University of Tokyo, Tokyo 113-0033,
Japan
\\
$^2$Department of Physics, Faculty of Science, University of Tokyo, Tokyo 113-0033,
Japan}

\abst{
We have investigated the relation of the direction of the momentum
among the matter, neutrino, and proto-neutron star in a
collapse-driven supernova in order to discuss the pulsar kick.
In particular,
we have investigated the effects of the pulsar motion on the explosion,
which are neglected in the previous study.
As a result, it is suggested that the direction of the total
momentum of the matter and neutrino is opposite to that of the
momentum of the proto-neutron star in the asymmetric
explosion models. This is because the center of the explosion deviates
from the center of the progenitor due to the pulsar motion. This
picture is common among the asymmetric explosion models.
So if we assume that the pulsar motion is caused by
an asymmetric supernova explosion, the neutron star born in SN 1987A,
which has not been found yet, will be moving in the southern part of the
remnant. In other words, if we can find one neutron star in SN 1987A
on the south part of the remnant, asymmetric explosion models will be
supported by the observation better than the binary models.
}

\begin{document}

\maketitle

\section{Introduction} \label{intro}
\indent

It is a well known fact that pulsars in our galaxy have velocities
much in excess of those of ordinary stars\cite{rf:hla93}.
It is reported that their
transverse speeds range from 0 to $\sim$ 1500 km $\rm s^{-1}$ and
their mean three-dimensional speed is 450 $\pm$ 90 km $\rm
s^{-1}$\cite{rf:ll94}.

On the other hand, there are many theoretical models in order to explain the
pulsar kick. One is that a neutron star in a binary system can escape
from the system with rapid speed due to a supernova explosion of the
nascent star\cite{rf:ggo70}. There are also many models in which
effects of asymmetric supernova explosion are taken into
consideration\cite{rf:bh96}\cite{rf:ks96}. For example, it is reported 
that a $1 \%$ anisotropic neutrino radiation from the proto-neutron star
results in a kick velocity consistent with the observations\cite{rf:ks96}.
However, there are few
observations to determine which model is the most promising one.

As for the asymmetric explosion models, only Burrows and Hayes (1996)
performed numerical simulations and estimated the speed of the
proto-neutron star in the supernova matter. 
They introduced the anisotropy of the system by artificially
decreasing the density of the Chandrasekhar core within 20$^{\circ}$
of the pole, which may be realized by the convection during the
stellar evolution. They also estimated the
contributions due to the neutrino emission anisotropy and the ejecta
motions. As a result, they reported that the proto-neutron star can be
accelerated to $\sim$ 530 km s$^{-1}$. They also concluded that the
direction of the total momentum of the matter and neutrino is opposite 
to that of the momentum of the proto-neutron star.

Recently, Nagataki (1999) discussed that the asymmetry of the observed
line profiles of $\rm Fe[II]$ in SN 1987A can be also explained by the
asymmetry of the explosion.
In the case of SN 1987A, more matter has to be conveyed to the north
side than to the south side in order to reproduce the observed
asymmetric line profiles.
As a result, if we believe the result of Burrows and Hayes (1996), the
proto-neutron star born in SN 1987A, which has not been found yet,will
be moving in the southern part of the remnant\cite{rf:n99}.

It is worth while discussing whether the conclusion of Burrows and
Hayes (1996) on the momentum of the matter, neutrino, and pulsar is
common among the asymmetric supernova explosion models. If so, the
conclusion on the location of the neutron star born in SN 1987A will
be valid as long as the asymmetric explosion models are believed.
Moreover, if the neutron star is discovered on the south side as
Nagataki (1999) predicted, the asymmetric
explosion models will be supported by the observation than the binary
models.

In this paper, we investigate whether the conclusion derived by Burrows and
Hayes (1996) on the momentum of the matter, neutrino, and pulsar is
common among the asymmetric supernova explosion models. In particular, 
we investigate the effects of the pulsar motion on the explosion,
which are neglected in the study of Burrows and Hayes (1996). 
In section~2, we show analytical estimates for the effects of the
neutrino heating by the proto-neutron star. In section~3, numerical
estimates of neutrino flux are shown. Discussion and conclusion are
presented in section~4.

\section{Analytical Estimates for the effects of the
neutrino heating} \label{analytic}
\indent

The mechanism of collapse-driven supernovae has been understood as follows
\cite{rf:b90}: when the mass of the iron core of the progenitor
exceeds the Chandrasekhar mass, the star begins
to collapse. The collapse continues until the central density of
the collapsing core reaches about (1.5-2) times the nuclear matter
density ($\rho = 2.7 \times 10^{14} \rm g\ cm^{-3}$),
beyond which matter becomes too stiff to be compressed further. A shock
wave then forms, propagates outward. At first, the shock wave is not so strong
and stall at $\sim 200$ km\cite{rf:bh96} in the iron core (it is called
a stalled shock wave). However,
by the continuous neutrino heating ($\sim 500$ ms\cite{rf:w85}), 
the shock wave is revived, begins to propagate outward again, 
and finally produces the supernova explosion.
This phenomenon is called as the delayed explosion, which is the most
promising theory for the mechanism of the collapse-driven supernova
explosion.

In this section, we investigate the contributions to the shock revival
due to the injection of the momentum and the thermal energy by the
neutrino heating, respectively.

\begin{figure}[htbp]
     \centerline{\epsfxsize=0.65\textwidth\epsfbox{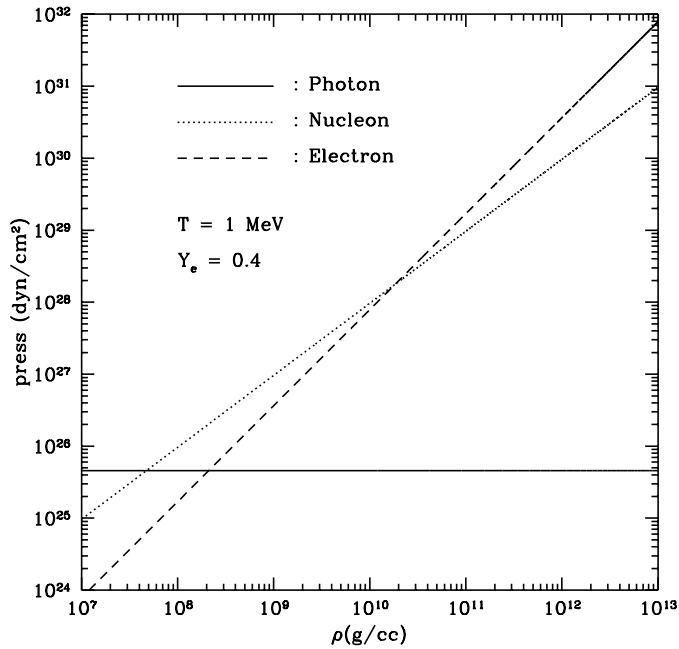}}
      \caption{
Density dependence of the pressure. Temperature and electron fraction
are set to be 1 MeV and 0.45. Solid line: the radiation pressure.
Dotted line: the ideal gas pressure. Short dashed line: the degenerate 
electron pressure.
}\label{fig1}
\end{figure}

Behind the stalled shock wave, the equation of state (EOS) is well
determined by the radiation pressure, the ideal gas pressure, and the
pressure of degenerate electrons at zero temperature. To put it
concretely, the pressure and the energy per unit volume are described as
follows: 
\begin{eqnarray}
    \label{eq:1}
p &=& a \it T{\rm ^4/3} + \it \rho k T/\mu m_u + p_e (\rho) \\
\rho \epsilon &=& a \it T{\rm ^4} + \rho k T / (\gamma_g \rm - 1) \it \mu m_u 
+ \rm 3 \it p_e (\rho)
\end{eqnarray}  
where $a$ is the radiation constant, $k$ Boltzmann constant, $m_u$ the 
atomic mass unit, $\mu$ the mean mass number of the gas, $\gamma_g$ the 
adiabatic index of the gas. The pressure of degenerate electrons $p_e$
is given by
\begin{eqnarray}
    \label{eq:2}
p_e &=& \rm 1.24 \times 10^{15} dyn \; cm^{-2}(\it Y_e \rho \rm )^{4/3}
\end{eqnarray}  
where $Y_e$ is the electron fraction of the system.
The typical values for the $\rho$, $T$, $\mu$, $\gamma_g$, and $Y_e$
behind the stalled shock wave
are\cite{rf:bhf95}\cite{rf:fh99}\cite{rf:jm95} $\rm
10^{10}$ g $\rm cm^{-3}$, 1 MeV, 1, 1.2, 0.4. The density dependence of the
pressure is shown in Fig.~\ref{fig1}.

At first we consider the contribution of the injection of the thermal
energy. From Eq.~(\ref{eq:1}) and (2.2), the partial
derivative of the pressure by the energy density can be written as follows: 
\begin{eqnarray}
    \label{eq:5}
\left. \frac{\partial p}{\partial \epsilon} \right|_{\rho} =  \left.
\frac{\partial p}{\partial T} \right|_{\rho} \left.
\frac{\partial T}{\partial \epsilon} \right|_{\rho} = \frac{\rho}{ 4
aT^3 + \frac{\rho k}{(\gamma_g - 1)\mu m_{u}}} \left(\frac{4}{3} aT^3 +
\frac{\rho k}{\mu m_{u}} \right). 
\end{eqnarray}
At the behind the stalled shock wave, the contribution of the
radiation can be neglected (see Fig.~\ref{fig1}). So we can rewrite
Eq.~(\ref{eq:5}) as
\begin{eqnarray}
    \label{eq:6}
\left. \frac{\partial p}{\partial \epsilon} \right|_{\rho} = 
(\gamma_g - 1) \rho \sim \frac{1}{5} \rho.
\end{eqnarray}
From Eq.~(\ref{eq:6}), we can estimate the thermal pressure by the
neutrino heating as
\begin{eqnarray}
    \label{eq:62}
p = \rho \epsilon /5.
\end{eqnarray}

Next, we consider the contribution of the injection of the momentum.
That is, we have to estimate the ram pressure which is the added pressure
when the flow is interrupted and the velocity becomes to be zero at
the stalled shock front. However, we can not use the Bernoulli's
equation behind the stalled shock wave. This is because the pressure
is determined not only by the electron but also by the ideal gas at
$\rho = 10^{10}$ g $\rm cm^{-3}$ (see Fig.~\ref{fig1}). In other words,
the pressure is not determined only by the density. So we have to give 
a rough order-estimation for the ram pressure, which will be enough
for the discussion in this section.
From the dimension analysis, we can estimate that the ram pressure is
as follows:
\begin{eqnarray}
    \label{eq:4}
p_{\rm ram} \sim \rho v^2
\end{eqnarray} 
where $p_{\rm ram}$ is the obtained ram pressure.

The typical explosion energy of a supernova is $\sim 10^{51}$
ergs\cite{rf:se88}. The total mass behind the stalled shock wave is
estimated to be $\sim 0.1 M_{\odot}$ when the average density behind
the stalled shock wave and the location of the stalled shock wave are
assumed to be $10^{10}$ g $\rm cm^{-3}$ and 200 km, respectively.
So the thermal energy density
behind the stalled shock wave is
estimated to be 5$\times 10^{18}$ erg $\rm g^{-1}$. So,
using Eq.~(\ref{eq:62}), the 
thermal pressure due to the neutrino heating is estimated to be  
$10^{28}$ dyn $\rm cm^{-2}$. On the other hand,
the velocity due to the momentum transfer from neutrinos
to the matter is estimated to be 2$\times 10^{8}$ cm $\rm s^{-1}$.
Using Eq.~(\ref{eq:4}), the ram pressure is estimated to be
4$\times 10^{26}$ dyn $\rm cm^{-2}$.

In this section, we find that the contribution to the pressure due
to the injection of the thermal energy will be greater than that due
to the injection of the momentum by the neutrino heating. However,
the ratio of the ram pressure relative to the thermal one ($\sim$ 1/25 
in our analysis) is small relative to their absolute value.
It will not be strange that their contributions become comparable in a
realistic calculation. So we will investigate both of their effects on
the explosion in the next section.

\section{Numerical estimates of Neutrino flux}
\subsection{Angular dependence of the normal component of the neutrino
flux}
As we noted in introduction, the proto-neutron star might have
moved from the center of the collapse due to the kick before the
explosion. If the dynamics of the explosion is controlled by
the thermal pressure of the matter heated by the emitted neutrino flux,
the integrated flux determines the magnitude of the explosion. In this
case it is expected that the neutron star (NS) heats the matter behind
the stoled shock wave anisotropically and the irons would not be blown
off spherically. In this subsection we show that there is an
anisotropy of the integrated neutrino flux received on the back of the
shock front and show how it depends on the azimuthal angel $\cos
\theta$ to the line of the kick.

For the simplification, we assume that the neutron star whose radius
is $=r_{ns}$ moves at a constant velocity $=v_{ns}$ along the line of
the kick from the center of the collapse. In addition we assume that
the neutron star starts to move as soon as it begins to emit the
electron neutrino and it keeps radiating the neutrino at a constant
flux $\F_0$ ($\erg \cm^{-2} \sec^{-1}$). Here let $t$ the passed time
since the neutron star started to move.  Then the distance from the
center of NS to the shock front is given by $r = l_0 - v_{ns}t$, where
$l_0$ is the radius of the shock front. The matter behind the stoled
shock wave receives the neutrino flux $\simeq (r_{ns}/r)^2 \F_0$ per
an unit time. Because NS has the rapid velocity enough to reach near
the shock front until the explosion, we consider the integrated flux
only for $0 \leq t \leq t_{max}$, where $t_{max}=(l_0-r_{ns})/v_{ns}$.

\begin{figure}[htbp]
        \centerline{\epsfxsize=0.65\textwidth\epsfbox{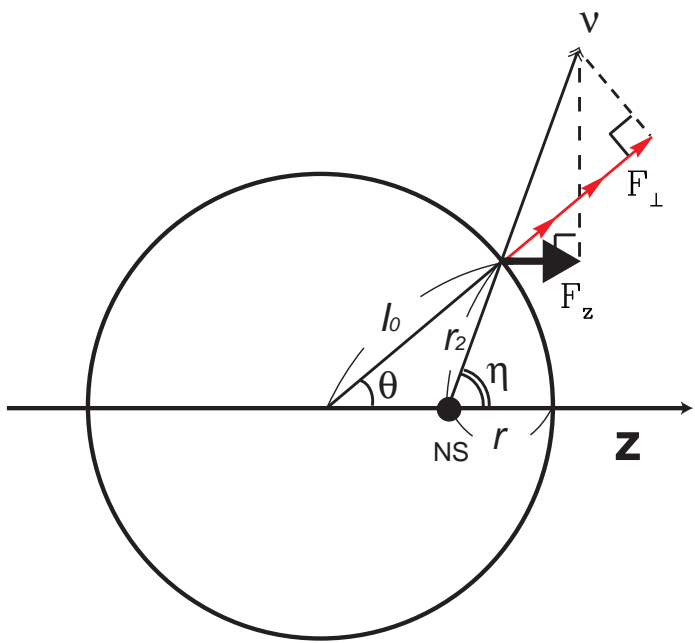}}
      \caption{
      Definition of coordinates and angles. The filled circle is the
      moving neutron star (NS). The outer circle represents the shock
      front. }
    \label{fig:angle}
\end{figure}

\begin{figure}[htbp]
     \centerline{\epsfxsize=0.65\textwidth\epsfbox{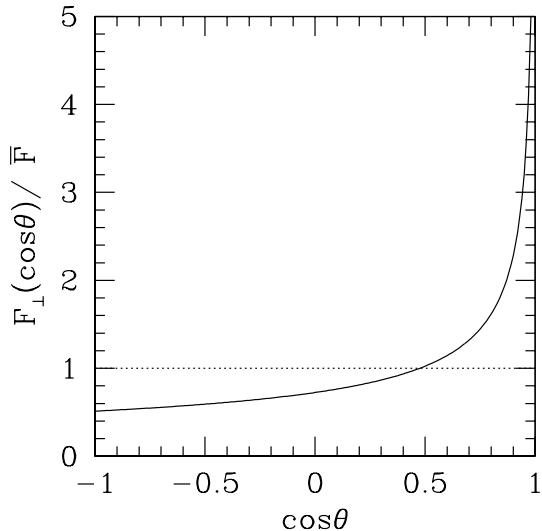}}
      \caption{
      Plot of the integrated normal component of the neutrino flux
      received on the back of the shock front as a function of the
      azimuthal angle $\cos \theta$. The integration of the flux is
      performed while NS is moving for $0 \leq t \leq t_{max}$. }
        \label{fig:cos_depend_perp}
\end{figure}

Here we take the line of the kick as z-axis. As we show in
Fig.~\ref{fig:angle}, let $r_2$ the distance from NS to an unit area
behind the shock front and $\eta$ the angle between the direction to
it and z-axis.
Then the angle $\eta$ is expressed by
\begin{equation}
    \label{eq:eta}
    \cos\eta = \frac{r - l_0(1-\cos\theta)}{r_2},
\end{equation}
where $r_2$ is given by $r_2^2=l_0^2+(l_0-r)^2-2l_0(l_0-r)\cos\theta$.
If the neutron star keeps radiating the neutrino until it reaches the
shock front, a net neutrino flux $\F_{\perp}(\cos\theta)$ received on
an unit area behind the shock front is expressed as the integrated
normal component of the neutrino flux for $0 \leq t \leq t_{max}$. Then
it is given by
\begin{eqnarray}
    \label{eq:int_neu_perp}
    \F_{\perp}(\cos\theta) &=& \int_0^{t_{max}} \F_0
                    \left(\frac{r_{ns}}{r_2}\right)^2 \cos(\eta-\theta)
                    dt \nonumber \\ 
             &=& \F_0 \frac{r_{ns}^2}{v_{ns}} 
                 \left ( \frac{l_0-r_{ns}}{l_o\sqrt{r_{ns}^2
                    +2l_0(1-\cos\theta)(l_0-r_{ns})}} \right).
\end{eqnarray} 
If the neutron star does not move at all, the normal component of the
integrated flux is equal anywhere behind the shock front and it is
given by
\begin{eqnarray}
    \label{eq:F_eq}
    \bar{\F} &=& \F_0 \left(\frac{r_{ns}}{l_0}\right)^2 \times t_{max}
             \nonumber \\ 
             &=& \frac{\F_0 (l_0-r_{ns})r_{ns}^2 }{v_{ns} l_0^2}.
\end{eqnarray}
Then the fraction of the received flux is estimated by
\begin{equation}
    \label{eq:frac_cos_perp}
    \frac{\F_{\perp}(\cos\theta)}{\bar{\F}} = \frac{l_0}{\sqrt{r_{ns}^2
    +2l_0(1-\cos\theta)(l_0-r_{ns})}}. 
\end{equation}
In Fig.~\ref{fig:cos_depend_perp} we plot
$\F_{\perp}(\cos\theta)/\bar{\F}$ as a function of $\cos\theta$. Here
we take the representative values $r_{ns}$= 10km, $l_0$=200 km and
$v_{ns}$=450 km sec$^{-1}$.

From Fig.~\ref{fig:cos_depend_perp} we find that if the explosion is
triggered by the thermal energy of the matter heated by the neutrino,
it is expected that for $\theta \ge \pi/3$ the matter is extremely
blown off. Namely, a small amount of the iron is strongly emitted
forward because of the large pressure gradient. On the other hand,
the majority is mildly pushed backward because the gradient of the
thermal pressure is smoother in the backward direction. 
In addition we can
see that the asymmetry of the received flux is much more larger than
1$\%$ which is needed for the initial kick of NS. 
It ensures that we can neglect the intrinsic anisotropy of the
neutrino emission from the proto-neutron star as we assumed.

\subsection{Angular dependence of the forward component of the
neutrino flux}
If the explosion is triggered by ram pressure caused by the momentum
transfer from the emitted neutrino to the matter just behind the
stoled shock wave, the magnitude of the explosion should be closely
related to the forward component of the neutrino flux. In this
subsection we show how the integrated forward (z-axis) component of
the neutrino flux depends on the azimuthal angle $\cos \theta$.

\begin{figure}[htbp]
     \centerline{\epsfxsize=0.65\textwidth\epsfbox{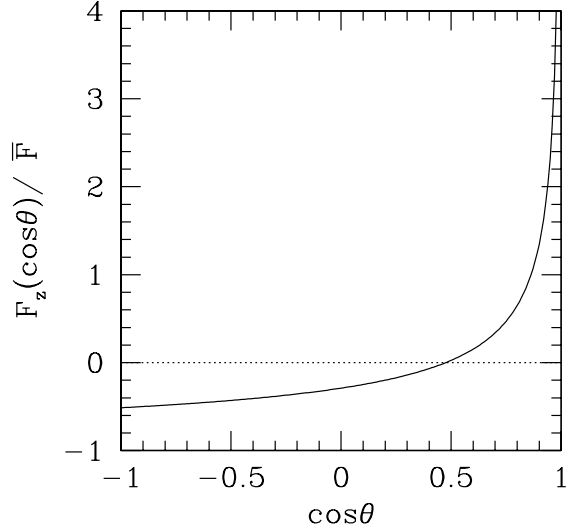}}
      \caption{
      Plot of the integrated forward component of the neutrino flux
      received on the back of the shock front as a function of the
      azimuthal angle $\cos \theta$. The integration of the flux is
      performed while NS is moving for $0 \leq t \leq t_{max}$.}
        \label{fig:cos_depend_z}
\end{figure}

If the neutron star keeps radiating the neutrino until it reaches the
shock front, the integrated forward component of the neutrino flux
received on an unit area is given by
\begin{eqnarray}
    \label{eq:int_neu_z}
    \F(\cos\theta)_z &=& \int_0^{t_{max}} \F_0
                    \left(\frac{r_{ns}}{r_2}\right)^2 \cos\eta dt \nonumber \\
             &=& \F_0 \frac{r_{ns}^2}{v_{ns}} 
                 \left ( \frac{1}{\sqrt{r_{ns}^2 +2l_0(1-\cos\theta)(l_0-r_{ns}
)}}-\frac1{l_0}\right).
\end{eqnarray} 
On the other hand, if the neutron star does not move at all, the
integrated flux is given in Eq.~(\ref{eq:F_eq}) as before.  Then the
fraction of the integrated forward component is estimated by
\begin{equation}
    \label{eq:frac_cos_z}
    \frac{\F_z(\cos\theta)}{\bar{\F}} = \frac{l_0^2}{l_0-r_{ns}}\left
    ( \frac{1}{\sqrt{r_{ns}^2
    +2l_0(1-\cos\theta)(l_0-r_{ns})}}-\frac1{l_0}\right). 
\end{equation}
In Fig.~\ref{fig:cos_depend_z} we plot $\F_z(\cos\theta)/\bar{\F}$ as
a function of $\cos\theta$. We find that if the explosion is caused by
the momentum transfer from the neutrino flux to the matter behind the
shock front, it blows off the matter forward only for $\theta \ge
\pi/3$ and the majority of the matter is pushed backward, which is
consistent with the conclusion derived in the previous subsection.
This means that our conclusion on the flow of the matter is valid
irrespective to the effects investigated.
Also,
because the asymmetry of the integrated neutrino flux is much larger
than 1$\%$ almost everywhere, we should not be worried about the inherent
asymmetric emission needed for the origin of kick. This means that
the fact that the large amount of the matter is
blown off backward does not depend on the
mechanism of asymmetric supernova explosion.

\section {Discussion and Conclusion} \label{conclusion}
\indent

We have discussed whether the conclusion derived by Burrows and
Hayes (1996) on the momentum of the matter, neutrino, and pulsar is
common among the asymmetric supernova explosion models. In particular,
we have investigated the effects of the pulsar motion on the explosion,
which are neglected in their study.
As a result, it was suggested by the present discussions that the
direction of the total
momentum of the matter and neutrino is opposite to that of the
momentum of the proto-neutron star in the asymmetric
explosion models. This is because the center of the explosion deviates
from the center of the progenitor due to the pulsar motion.
This picture is common among the asymmetric explosion models.
So if we assume that the pulsar motion is caused by
an asymmetric supernova explosion, the neutron star in SN 1987A, which
has not been found yet, will be moving in the southern part of the
remnant.

Moreover, the discovery of the mystery spot,
separated from SN 1987A to the south region by 60 mas\cite{rf:ne87},
will support our conclusion. The existence of the bright spot on the
south region will
suggest that the existence of the strong shock wave on the south
region, which is consistent with our conclusion (see
Fig.~\ref{fig:cos_depend_perp} and~\ref{fig:cos_depend_z}). 
Here we have to comment on the second bright source which was reported
to be detected on the north side of the remnant\cite{rf:np99}.
In their paper, it was concluded that the south spot has to be
red-shifted and the north side spot has to be blue-shifted. It means that
the south part of the ring around SN 1987A is nearer to us than the
north part if these bright spots are ejected from the polar regions.
This conclusion is opposite to the ones derived by the other many
observations on the ring\cite{rf:pla95}\cite{rf:mic98}.
In this study, we have assumed that the north side of the ring is nearer
to us and concluded that the neutron star born in SN 1987A is running
in the south region of the remnant. However, if we believe the
conclusion derived from the study of the second bright spot, our
conclusion on the location of the neutron star is changed oppositely.
We also give an additional comment on the mechanism of the jet-induced
explosion which may explain the existence of the mystery
spot\cite{rf:nak87}\cite{rf:kho99}\cite{rf:mw99}. These are the
promising ones which
may explain the existence of the mystery spot, but there has to be
a mechanism which breaks the symmetry with respect to the equatorial plane.
This is because the mystery spot seems to exist on only one side
as discussed above (at least, the second bright source on the north side is
much fainter than the one on the south side). The effects of the
pulsar motion on the asymmetric explosion discussed in this study
explain such an asymmetry with respect to the equatorial plane naturally.

As mentioned above, the asymmetric explosion models can explain
naturally the asymmetry of the iron line profiles and the existence of
the mystery spot at the same time. Also it is suggested that the
pulsar born in SN
1987A is running in the south part of the remnant. On the other
hand, the simple binary models in which the explosion is assumed to be
spherically symmetric can not explain these observations and predicts
that the pulsar is located at the center of the remnant.
Moreover, the only possible candidate as a companion of the
progenitor of SN 1987A is a compact object
(neutron star or black hole)\cite{rf:cfb99}. In this case, two compact
objects will be found in the remnant of SN 1987A. One of them is found
at the center of the remnant, if we believe the simple binary models.
In other words, if we can find one neutron star in SN 1987A
on the south part of the remnant and deny the existence of the another
compact object at the center of the remnant, asymmetric explosion
models will be supported by the observation better than the binary
models.

There may be a possibility to detect anisotropy in the young supernova 
ejecta using VLBA as long as it is located within $\sim$ 100
Mpc\cite{rf:p99}. There may also be a possibility to find a pulsar in
the ejecta. Increase of such observations will make it clear the
relation between the asymmetric explosion and pulsar kick. 

In the delayed explosion model, about 1$\%$ of the
neutrino energy is transfered into the energy of the explosion. In
this situation we should delicately treat the neutrino
transfer, and we should discuss such a subtle problem carefully. However,
to qualitatively understand the physical mechanism in the model, we think
that the neutrino flux and its integrated value give the adequate
informations. Therefore in this study, we have
investigated the effects of the pulsar motion on the explosion using
simple analysis. It will be necessary to check our conclusion by
performing realistic numerical simulations in which the effects of
neutrino heating and its back reaction are taken into consideration.
In the present circumstances, the effects of the pulsar motion in the
iron core have not been taken into account and we have treated the
neutrino flux as a simple thermal radiation in the numerical
simulations concerning with the collapse-driven supernova explosion.
However, we can understand its importance easily because the neutron
star can reach to the stalled shock front in $\sim$ 500 ms, which is
the dynamical timescale of the stalled shock wave, as long as the
neutron star moves with the observed mean speed (450 km $\rm s^{-1}$).
Even if we assume that the pulsar is accelerated constantly and its
velocity becomes to be 450 km $\rm s^{-1}$ in 500 ms, the location of
the pulsar at $t$ = 500 ms becomes to be $\sim$ 110 km from the
center. In fact, Burrows and Hayes (1996) reported that the pulsar
gets the velocity $\sim$ 500 km $\rm s^{-1}$ in 200 ms. We can easily
guess that the effects of the pulsar motion on the dynamics of the
explosion should be taken into consideration.  Such calculations are
now underway and we will report the results in the near future.

\section*{Acknowledgements}
We would like to thank Dr. S. Yamada for useful discussions.
This research has been supported in part by a Grant-in-Aid for the
Center-of-Excellence (COE) Research (07CE2002) and for the Scientific
Research Fund (199804502, 199908802) of the Ministry of Education,
Science, Sports and Culture in Japan and by Japan Society for the
Promotion of Science Postdoctoral Fellowships for Research Abroad.


\end{document}